\documentclass[12pt]{iopart}
\usepackage{setstack,iopams,bm}
\usepackage{graphicx}

\def\bs{\hspace{-5pt}}
\def\II{\mathcal{I}}
\def\DD{\mathsf{D}}
\def\text#1{{\rm #1}}
\def\eqref#1{(\ref{#1})}
\def\what#1{\widehat{#1}}
\def\notag{\nonumber}
\usepackage{hyperref}

\begin{document}

\title[Accelerating small-angle scattering experiments with machine learning]{Accelerating small-angle scattering experiments with simulation-based machine learning}

\author{Takuya Kanazawa, Akinori Asahara and Hidekazu Morita}
\address{Hitachi, Ltd., Tokyo 100-8280, Japan}
\ead{takuya.kanazawa.cz@hitachi.com}

\begin{abstract}
Making material experiments more efficient is a high priority for materials scientists who seek to discover new materials with desirable properties. In this paper, we investigate how to optimize the laborious sequential measurements of materials properties with data-driven methods, taking the small-angle neutron scattering (SANS) experiment as a test case. We propose two methods for optimizing sequential data sampling. These methods iteratively suggest the best target for the next measurement by performing a statistical analysis of the already acquired data, so that maximal information is gained at each step of an experiment. We conducted numerical simulations of SANS experiments for virtual materials and confirmed that the proposed methods significantly outperform baselines. 
\end{abstract}

\submitto{J. Phys.: Mater.}
\maketitle

\section{Introduction}

In recent years, Materials Informatics (MI) has gained popularity as a data-driven approach to materials discovery and design \cite{Rajan2005,LookmanBook2016,Ramprasad2017,Liu2017,Butler2018,Gubernatis2018,Schleder2019}. While the success of MI hinges on materials datasets with wide coverage and high accuracy, such datasets are not yet sufficiently available in many fields of physics and chemistry. Thus the demand for high-cost manual experiments to garner new data remains high. Precise determination of material properties often requires a sequence of elaborate measurements, and it is highly desired to expedite such experiments in order to reduce the time, money, and effort involved. Thus far, the application of machine learning techniques in MI to the optimization problem of such serial measurements has been rather limited. Even though the use of Bayesian optimization to design sequential material experiments is common \cite{Seko2015,Frazier2015,Ueno2016,Lookman2017}, the typical goal of such studies is to discover a material with the best property as quickly as possible, in contrast to sequential measurements on a single material sample, where the challenge is to accurately determine material properties by optimally integrating multiple measurement outcomes. 

In this paper, we explore optimization strategies for sequential experiments in materials science, taking the small-angle neutron scattering (SANS) experiment \cite{Feigin1987,Jaksch2019} as a representative example. SANS is widely used to study the microstructures of various materials including alloys, ceramics, and polymers, but the design of an optimal SANS experiment has been left to manual adjustment by experienced researchers. To alleviate this burden, we formulate the design of a SANS experiment as a multi-step decision making problem and introduce two methods that enable adaptive measurement planning. The core idea is to create a database of virtual experiments by means of simulation and then use it adaptively during the real experiment. In the first method, we use simulation to fix the best sampling procedure for each of many virtual samples and record the procedures in a database. During the actual experiment, we send the acquired data into a similarity search over the database and retrieve the best sampling protocol for the most similar sample. In the second method, the outcomes of future measurements during an experiment are predicted in two distinct ways: one uses the database of simulated measurements, while the other does not. The target of the next measurement is then determined through comparison of the two predictions. 

To benchmark the proposed methods, we prepared 100 virtual sample materials and simulated SANS experiments on them. The results showed that the experimental duration required to achieve satisfactory accuracy was reduced by 50--65\% with these methods, as compared to random sampling with no prior planning. We thus conclude that SANS experiments can be accelerated by a factor of 2--3 with our method. Since the small-angle X-ray scattering (SAXS) \cite{Feigin1987,Jaksch2019} also shares the same principles with SANS, we believe our methods will work for SAXS experiments as well.

In section~\ref{sc:setup} of this paper, we summarize the setup of the problem and provide a pedagogical toy example to highlight the methods we propose. In section~\ref{sc:sansdefs}, an overview of a SANS experiment is given and relevant terminology is introduced. In section~\ref{sc:methods}, new methods for planning an optimal SANS experiment are defined and discussed in detail. In section \ref{sc:exp}, these methods are tested in numerical simulations and compared with baseline methods. We conclude in section~\ref{sc:conc} with a brief summary and mention of future work.

\section{Problem Setup\label{sc:setup}}

Let us consider a class of problems characterized by a relation of the form
\begin{equation}
	\mathbf{y} = {\cal F}(\mathbf{x}) + \mathbf{n},
\end{equation}
where $\mathbf{y}$ is an observed signal, $\mathbf{x}$ is a hidden property that we want to uncover, $\mathbf{n}$ is noise, and ${\cal F}(\cdot)$ is a function of known form. The task is to reconstruct $\mathbf{x}$ from a partial observation of $\mathbf{y}$. This class of problems includes a number of interesting applications such as image denoising, where $\mathbf{x}$ is a clean image and $\mathbf{y}$ is a corrupted image, and compressed sensing, where ${\cal F}$ is a linear map and $\mathbf{x}$ is assumed to be sparse. SANS experiments belong to this class, as explained in section~\ref{sc:sansdefs}. 

The question we examine is how to find a policy for sequentially measuring the components of $\mathbf{y}$ so as to maximize the efficiency of recovering $\mathbf{x}$. To illustrate key issues clearly, let us consider a toy problem. Suppose we wish to infer a periodic function $f(x)$ defined on the interval $x\in[-\pi,\pi]$. The allowed operation is to query the value of the integral, 
\begin{equation}
	\alpha_k(f):=\frac{1}{2\pi}\int_{-\pi}^{\pi} \rmd x\;f(x) \,\rme^{-ikx} \qquad \text{for}~k\in\mathbb{Z}\,,
	\label{eq:alphadef}
\end{equation}
which we call a \emph{measurement} of $f$. After a set of measurements $\mathcal{K}_n=\{k_1,k_2,\cdots,k_n\}$, we may guess $f$ as $\what{f_n}(x)=\sum_{k\in\mathcal{K}_n} \alpha_{k}(f) \rme^{ikx}$. The discrepancy from true $f$ may be quantified by the mean squared error: 
\begin{equation}
	\delta_n:=\frac{1}{2\pi}\int_{-\pi}^{\pi}\rmd x\;|f(x)-\what{f_n}(x)|^2=\sum_{k\not \in \mathcal{K}_n}|\alpha_k(f)|^2\,. 
	\label{eq:delta_n}
\end{equation}
When a measurement is either costly or time-consuming, we want $\delta_n$ to decrease with $n$ as fast as possible so that the measurement can be truncated early. As is evident from \eqref{eq:delta_n}, the best strategy is then to sequentially measure $\alpha_k$ in descending order of magnitude, but this is impossible without knowing $f$ in advance. The challenge is to judge which $\alpha_{k_{n+1}}$ to measure next on the basis of previous measurements $\{\alpha_{k_1},\cdots,\alpha_{k_n}\}$. This is a nontrivial multi-step decision making problem. It is crucial to take into account correlations of $\{\alpha_k\}$ properly. 

We may conceive of two qualitatively different ways to tackle this problem: one is a deductive approach and the other is an inductive approach. They are outlined as follows. 
\begin{itemize}
\item {\bf Method 1 \text{(deductive)}.} To begin with, we \emph{assume} that $f$ is generated from the probability distribution $p(g)\propto \exp\left(-\lambda\int_{-\pi}^{\pi}\rmd x\;|\nabla g(x)|^2\right)$ over smooth periodic functions on $[-\pi,\pi]$, with $\lambda>0$ a parameter to control the smoothness. This is a working hypothesis posed to model the correlations among $\{\alpha_k\}$. In the Bayesian sense, $p(g)$ represents the prior distribution of $f$, i.e., the belief we have for $f$ before any measurement. After the measurements $\mathcal{K}_n$, the posterior probability $p(g|\mathcal{K}_n)$ can be factored as $p(\mathcal{K}_n|g)p(g)$ up to an irrelevant constant, based on the Bayes rule \cite{BishopBook}. We proceed further to modeling $p(\mathcal{K}_n|g)$ by a Gaussian distribution as 
\begin{equation}
	p(\mathcal{K}_n|g) \propto \exp\left(-\frac{1}{\sigma^2}\sum_{k\in\mathcal{K}_n} \left|\alpha_k(f) - \frac{1}{2\pi}\int_{-\pi}^{\pi}\rmd x\;g(x)\,\rme^{-ikx}\right|^2\right)\,.
\end{equation}
Combining this with the prior distribution, we find that the \emph{maximum a posteriori (MAP) estimation} \cite{BishopBook} of $f$ is given by
\begin{equation}
	\fl
	\what{f_n}^{\rm MAP} = \underset{g}{\rm arg\,min}\left(
	\frac{1}{\sigma^2}\sum_{k\in\mathcal{K}_n} \left|\alpha_k(f) - \frac{1}{2\pi}\int_{-\pi}^{\pi}\rmd x\;g(x)\,\rme^{-ikx}\right|^2
	+ \lambda \int_{-\pi}^{\pi}\rmd x\; |\nabla g(x)|^2 \right).
	\label{eq:fmap}
\end{equation}
In practice, this can be solved with numerical optimization algorithms. It is then straightforward to expand $\what{f_n}^{\rm MAP}$ into a Fourier series. The target of the next measurement is given by
\begin{equation}
	k_{n+1} = \underset{k\not \in \mathcal{K}_n}{\rm arg\,max}~\Big|\alpha_k\Big(\what{f_n}^{\rm MAP}\Big)\Big|\,.
\end{equation}
The essence of this method lies in guessing the outcome of future measurements ($\alpha_k(f)$ for $k\not\in\mathcal{K}_n$) from previous measurements ($\alpha_k(f)$ for $k\in\mathcal{K}_n$) by using an analytical Ansatz for the distribution of $f$.
\item {\bf Method 2 \text{(inductive)}.} We start by creating a database of periodic functions on $[-\pi,\pi]$ and their Fourier coefficients,
\begin{equation}
	\DD = \Big\{\Big(g_\ell, \{\alpha_k(g_\ell) \}_{k\in\mathbb{Z}}\Big)\Big\}_{\ell=1}^{N}\,.
\end{equation}
The data-generating process behind $\{g_\ell\}_\ell$ is supposed to be the same or approximately the same as the one that generated $f$. After the first $n$ measurements $\mathcal{K}_n$, we compare the results $\{\alpha_{k}(f)|k\in\mathcal{K}_n\}$ with those stored in the database $\{\alpha_{k}(g_\ell)|k\in\mathcal{K}_n\}$ and pick out the $g_\ell$ that most resembles $f$. Formally stated, we choose 
\begin{equation}
	g_\ell = \underset{g\in\DD}{\rm arg\,min}~\sum_{k\in\mathcal{K}_n}|\alpha_k(f)-\alpha_k(g)|\,,
	\label{eq:gellchoice}
\end{equation}
where the similarity is measured by the $L_1$ norm, but other choices are also possible. Then we decide on the target of the next measurement as
\begin{equation}
	k_{n+1} = \underset{k\not \in \mathcal{K}_n}{\rm arg\,max}~|\alpha_k(g_\ell)| \,.
\end{equation}
Just like method 1, method 2 also guesses the outcome of future measurements ($\alpha_k(f)$ for $k\not\in\mathcal{K}_n$) from previous measurements ($\alpha_k(f)$ for $k\in\mathcal{K}_n$), but this time, instead of making a parametric Ansatz for the distribution of $f$, we rely on a database and make a \emph{purely empirical} guess, in a manner analogous to the nearest neighbor algorithm in machine learning \cite{BishopBook}. 
\end{itemize}

In the above, we outlined two contrasting methods for measurement planning for the simple toy problem. As described in the following sections, the SANS experiment in materials science is technically much more involved than this example, but the basic ideas underlying methods 1 and 2 above can be carried over. Specifically, method 1 for SANS (section~\ref{sc:m1}) is purely empirical and is akin to method 2 for the toy problem, whereas method 2 for SANS (section~\ref{sc:m2}) is a hybrid of inductive and deductive methods. Why is a hybrid approach useful? To answer this, we modify the setup of the toy problem slightly as follows. Below \eqref{eq:alphadef} the estimate of $f$ based on the measurements $\mathcal{K}_n$ was given by $\what{f_n}(x)=\sum_{k\in\mathcal{K}_n} \alpha_{k}(f) \rme^{ikx}$. It amounts to simply dropping the contributions from all the unmeasured Fourier coefficients ($\alpha_k(f)$ for $k\not\in\mathcal{K}_n$). Actually, we can make a better guess of $f$ by making use of the MAP estimate \eqref{eq:fmap} to infer unmeasured coefficients, which results in the estimate
\begin{equation}
	\what{f_n}(x) = \sum_{k\in\mathcal{K}_n} \alpha_{k}(f) \rme^{ikx} 
	+ \sum_{k\not \in\mathcal{K}_n} \alpha_{k}\Big(\what{f_n}^{\rm MAP}\Big) \rme^{ikx}\;.
\end{equation}
This yields the discrepancy
\begin{equation}
	\delta_n=\frac{1}{2\pi}\int_{-\pi}^{\pi}\rmd x\;|f(x)-\what{f_n}(x)|^2
	=\sum_{k\not \in \mathcal{K}_n}\Big|\alpha_k(f)-\alpha_{k}\Big(\what{f_n}^{\rm MAP}\Big)\Big|^2 \,,
\end{equation}
where the right-hand side has been modified from \eqref{eq:delta_n}. An ideal choice for $k_{n+1}$ would be the $k$ for which $\Big|\alpha_k(f)-\alpha_{k}\Big(\what{f_n}^{\rm MAP}\Big)\Big|$ is largest, or in other words, the $k$ for which the MAP estimate of $\alpha_k$ is most inaccurate. However, such selection of $k_{n+1}$ is not feasible when $\alpha_k(f)$ for $k\not\in\mathcal{K}_n$ are unknown. To bypass this difficulty we return to method 2 above. Namely, we create a catalogue of periodic functions and their Fourier coefficients, and pick out a function $g_\ell$ that best matches the measured coefficients, as in \eqref{eq:gellchoice}. The next target of the measurement is decided following
\begin{equation}
	k_{n+1} = \underset{k\not\in\mathcal{K}_n}{\rm arg\,max}\; 
	\Big|\alpha_k(g_\ell)-\alpha_{k}\Big(\what{f_n}^{\rm MAP}\Big)\Big|\,.
\end{equation}
This method, combining the MAP estimate and the empirical database method, can be adapted to a SANS experiment very effectively, as described later in section~\ref{sc:m2}.

\section{Background and definitions\label{sc:sansdefs}}

\begin{figure}[bt]
	\centering
	\includegraphics[width=.5\columnwidth]{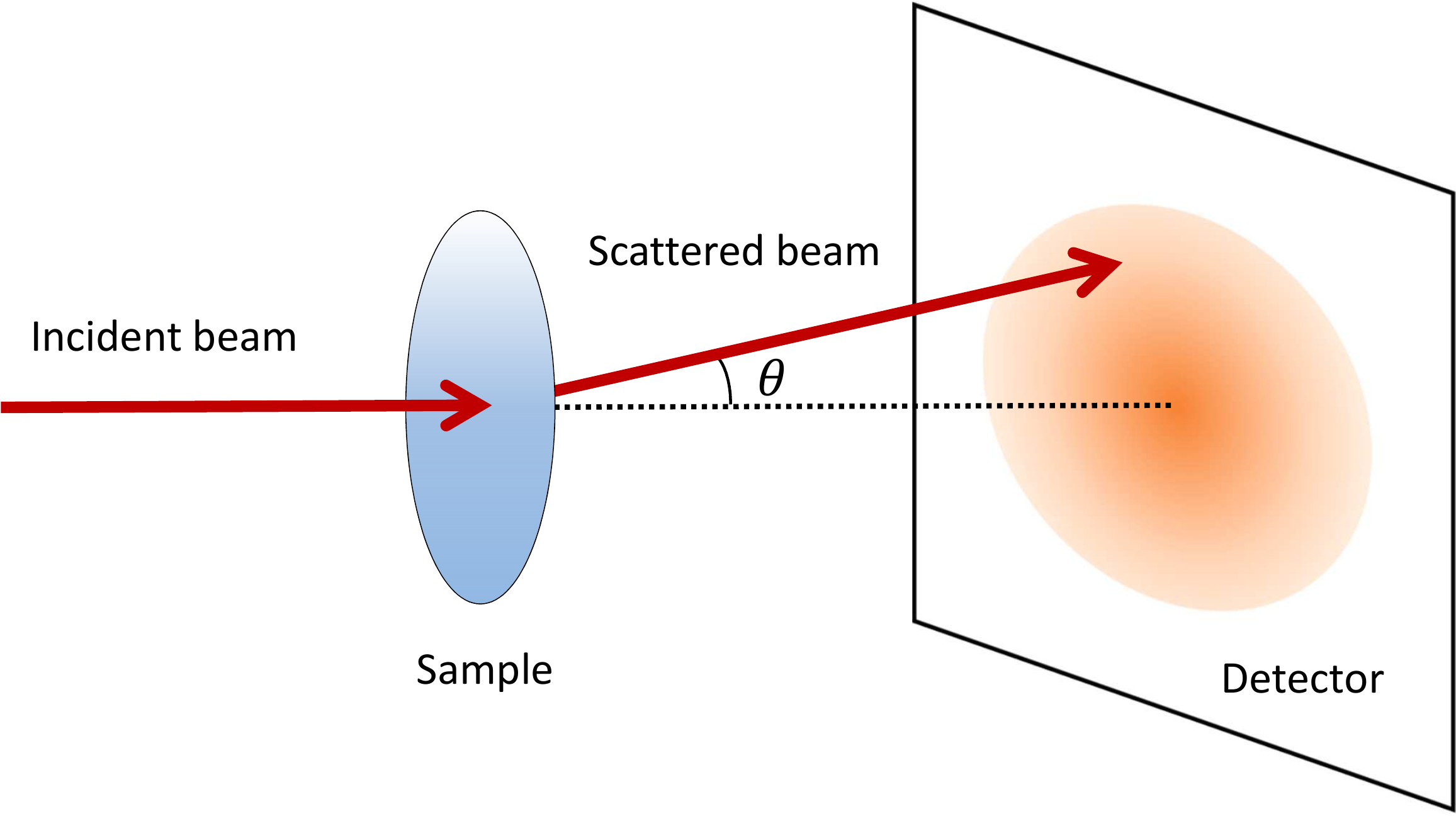}
	\caption{\label{fg:sans}Small-angle neutron scattering experiment.}
\end{figure}

In this section we provide a minimal overview of a SANS experiment and related data analysis steps, referring the reader to \cite{Feigin1987,Jaksch2019} for further details. SANS is an experimental technique to investigate the structures of various substances at a scale of about 1--100 nm. The setup of a typical SANS experiment is depicted in Figure~\ref{fg:sans}. 
A beam of neutrons is directed at a sample, and neutrons scattered at small angles are captured at a detector. The obtained signal is called the \emph{neutron scattering intensity} and is denoted by $I(q)$, where $q$ is the magnitude of the scattering vector. It is related to the scattering angle $\theta$ as $q=\frac{4\pi}{\lambda}\sin(\theta/2)$, with $\lambda$ the wavelength of the neutrons. $I(q)$ is proportional to the absolute square of a \emph{form factor} reflecting the geometry of the scatterers in the sample. If a scatterer is a homogeneous sphere of radius $r$, the form factor is given by \cite{Jaksch2019}
\begin{equation}
	F(q, r) = \frac{\sin(qr)-qr\cos(qr)}{(qr)^3}\,.
\end{equation}
If the ensemble of scatterers consists of spheres of various sizes following a probability distribution $p(r)$, then we have
\begin{equation}
	I(q) = \int_0^\infty \bs \text{d}r\; p(r) F(q, r)^2 
	\label{eq:Idef}
\end{equation}
where the overall multiplicative factor on the right hand side has been set to unity for simplicity of exposition (see \cite{Jaksch2019} for full details). 
After performing an experiment, we acquire the SANS data $\{I(q_n)\}_{n=1}^{N}$. The purpose of the experiment is to estimate the distribution $p(r)$ from this data. Unfortunately, this is an ill-posed inverse problem, for it is impossible to uniquely specify a continuous function from a finite number of data points. A traditional approach to obtain a reasonable $p(r)$ is the indirect Fourier transform (IFT) \cite{Glatter1977,Moore1980,Hansen1991,Muthig2016}. In the IFT approach, $p(r)$ is expanded by a set of basis functions as
\begin{equation}
	p(r) = \sum_{k}^{}a_k B_k(r)\,. \label{eq:p_exp}
\end{equation}
This expansion is plugged into \eqref{eq:Idef} to obtain the scattering intensity $\widehat{I}(q)$. The coefficients $\{a_k\}$ are then fixed in such a way that $\widehat{I}(q)$ closely approximates the observed data:
\begin{equation}
	\{a_k\} = \underset{a}{\rm arg\,min}\Big[
		\sum_{n=1}^{N}
		\big\{ I(q_n) - \widehat{I}(q_n) \big\}^2 + \lambda \Omega(a)
	\Big]\,.
	\label{eq:akinfer}
\end{equation}
Here, $\Omega(a)$ is a penalty term controlling the smoothness of the solution, and $\lambda$ is a coefficient for tuning the relative importance of the penalty. We note the parallel between \eqref{eq:fmap} and \eqref{eq:akinfer}. 

The obtained $\{a_k\}$ are plugged into \eqref{eq:p_exp} to give the estimated size distribution $\widehat{p}_N(r)$ based on the $N$ observed data points. As $N$ grows, $\widehat{p}_N(r)$ is anticipated to converge to the true distribution $p(r)$. To monitor the convergence, we measure the discrepancy of the two functions by the $L_1$ distance
\begin{equation}
	\mathcal{D}^{\rm L1}(p,\widehat{p}_N) 
	= \int_0^\infty \bs \text{d}r\; |p(r)-\widehat{p}_N(r)|, \label{eq:l1}
\end{equation}
and the Kullback--Leibler divergence
\begin{equation}
	\mathcal{D}^{\rm KL}(p,\widehat{p}_N) 
	= \int_0^\infty \bs \text{d}r\; p(r) \ln \frac{p(r)}{\widehat{p}_N(r)} \,. 
	\label{eq:KL}
\end{equation} 
These are the analogs of \eqref{eq:delta_n} in the toy problem. 

The goal of this work is to figure out how to find a sequence of measurements $\{q_1, q_2,\cdots,q_N\}$ that is optimal in the sense that the discrepancies \eqref{eq:l1} and \eqref{eq:KL} decrease as fast as possible. In the next section we propose two methods that adaptively suggest the next measurement $q_{n+1}$ according to the previous measurement results $\{I(q_1), \cdots, I(q_n)\}$. Throughout this work, for simplicity, we neglect the noise associated with measurements. In practice the noise can be suppressed by extending the measurement duration to reduce the statistical fluctuation of neutron counts at the detector.

\section{Proposed Methods\label{sc:methods}}

\subsection{Method 1:~Measurement Protocol Retrieval Based on Real-Time Similarity Search\label{sc:m1}}

This method is a special kind of \emph{case-based planning} \cite{Spalzzi2001}. The main idea of case-based planning is to reuse past successful plans in order to solve new planning problems \cite{Spalzzi2001}. We adapt this idea to the problem of SANS experiments as follows: we generate many virtual materials and create an optimal measurement plan for each of them via simulation. These plans are saved in a database, together with the profiles of respective materials. In a real SANS experiment, we match the current situation with those in the database and select the plan that best matches the current situation. The next measurement is carried out according to this plan. This procedure is repeated after every measurement until the experiment is terminated. Notably, this method requires preparation of a sufficiently large database before an actual experiment begins. How large it should be depends on the diversity of materials on which SANS experiments are performed.  

Let us describe the details. This method starts by stochastically creating many virtual sample materials with a variety of size distributions $\{p_k(r)\}_{k=1}^K$, where $K$ denotes the number of virtual samples. Let $N$ denote the maximum number of measurements per sample. For each sample, an ``optimal'' measurement plan $Q=\{q_1,\cdots,q_N\}$ is generated (we shortly describe how to do this), and the scattering intensities $\{I(q_1), \cdots, I(q_N)\}$ are computed via \eqref{eq:Idef}. Next, we create a database
\begin{equation}
	\DD = \Big\{ \Big[ \{I_k(q_n)\}_{n=1}^{N}, Q_k \Big] \Big\}_{k=1}^{K}
	\label{eq:DD}
\end{equation}
where the pair of the scattering intensity and the best plan is stored for every sample. This completes the first step of Method 1.

\begin{figure}
	\centering
	\includegraphics[width=.5\columnwidth]{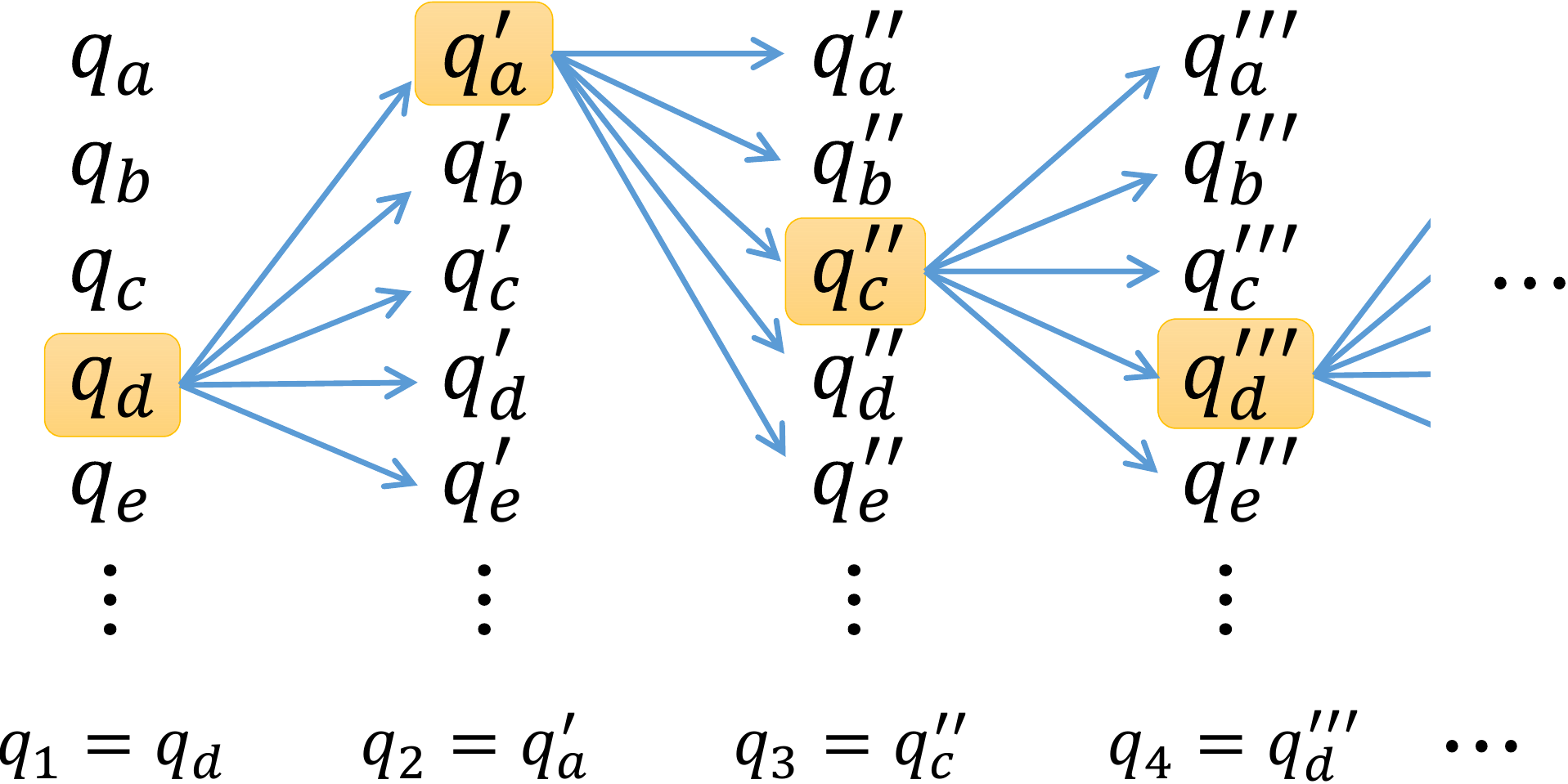}
	\caption{\label{fg:tt}Greedy optimization of the measurement plan.}
\end{figure}

An optimal plan $Q_k$ for a sample $k$ can be generated incrementally as depicted in Figure~\ref{fg:tt}. 
Let $1\leq n \leq N-1$. Given $q_1,\cdots,q_n$, we randomly create many $q$ as candidates for $q_{n+1}$. 
\footnote{We have generated a candidate $q_{n+1}$ randomly from a log-uniform distribution over a range $[q_{\rm min}, q_{\rm max}]$. Hence  the $N\cdot K$ values of $q_n$ that appear in $\DD$ are likely all distinct. The precise values of $q_{\rm min}$ and $q_{\rm max}$ used in our numerical experiment will be specified in section~\ref{sc:expsettings}.}
For each candidate $q_{n+1}$, we compute $\widehat{p}_{n+1}$ via the IFT [recall \eqref{eq:p_exp} and \eqref{eq:akinfer}] and estimate the discrepancy $\mathcal{D}(p_k, \widehat{p}_{n+1})$. We then choose the $q$ giving the smallest value of $\mathcal{D}(p_k, \widehat{p}_{n+1})$ as $q_{n+1}$. This process is iteratively repeated for $n=1,2,\cdots,N-1$ until the entire plan $(q_1,\cdots,q_N)$ is obtained. 

\begin{figure}
	\centering
	\includegraphics[width=.75\columnwidth]{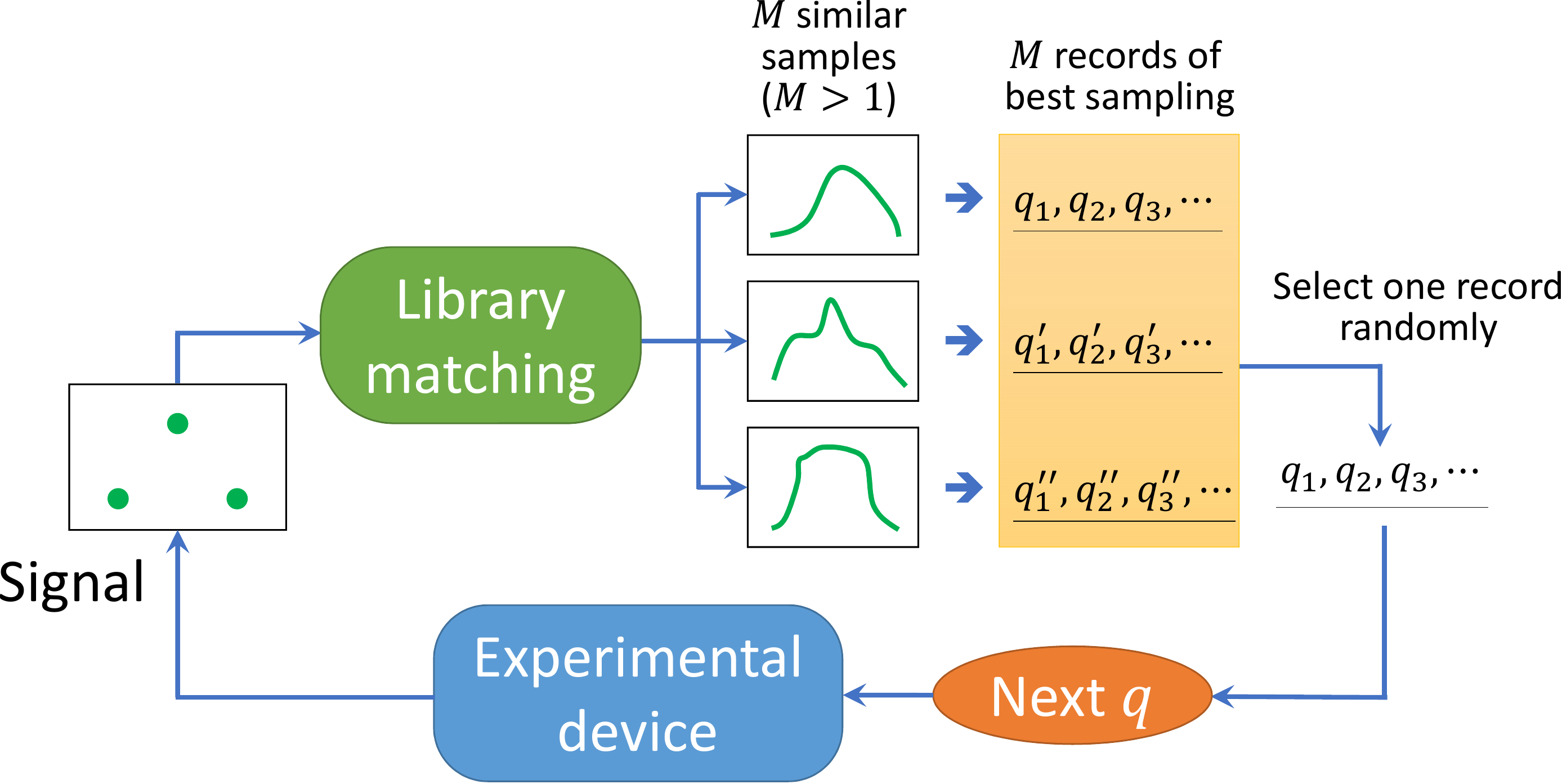}
	\caption{\label{fg:me1}Summary of Method 1.}
\end{figure}

When we have a new sample and want to conduct new measurements on it, the ``best'' plans stored in $\DD$ cannot be used immediately, because the plan $Q_k$ for a specific sample $k$ is not necessarily suitable for a new sample. The second step of Method 1 fills this gap. Suppose that the data $\{I(q_1),\cdots,I(q_n)\}$ is obtained from measurements on a new sample. We then conduct a similarity search on $\DD$ to find $M$ samples whose scattering intensities most resemble the observed data. This allows us to retrieve the $M$ best plans $\{Q_1,\cdots,Q_M\}$ accordingly. Next, we randomly select one plan, say $\widehat{Q}=(\hat{q}_1,\cdots,\hat{q}_N)\in\{Q_1,\cdots,Q_M\}$, and construct a subset $\widehat{Q}'=\{\hat{q}_i\in \widehat{Q}\,|\,\underset{1\leq j\leq n}{\min}|\hat{q}_i-q_j|>\delta\}$, where $\delta$ is a small positive constant. This means that all $\hat{q}_i$ that are too close to the already measured $q$ are left out to avoid nearly duplicate measurements. Finally, we select the $\hat{q}_i\in\widehat{Q}'$ with the smallest index $i$ as the next measurement point, $q_{n+1}$. This process is iterated until $N$ measurements $(q_1,\cdots,q_N)$ have been performed. The set of $M$ samples is updated after every new measurement. This completes the second step of Method 1. Figure~\ref{fg:me1} schematically illustrates the overall procedure. 

It is intuitively rather clear that $\{Q_1,\cdots,Q_M\}$ should serve as good plans for the new sample, as the observed data indicates that the new sample is similar to those $M$ samples in $\DD$. More nontrivial is the fact that letting $M>1$ helps to avoid overfitting. This is analogous to the situation in the multi-armed bandit problem \cite{Sutton2018Book}, which requires balancing exploration and exploitation. Setting $M=1$ is akin to pure exploitation, whereas $M>1$ allows for exploration. Empirically, we found that $M=3$ was most effective, and increasing $M$ further did not bring improvement. 

In Method 1 it is essential that the similarity search over $\DD$ works accurately. This hinges on how the similarity of two scattering data, $I_1(q)$ and $I_2(q)$, is defined. A naive definition such as $\sum_{i}|I_1(q_i)-I_2(q_i)|$ fails severely, because the scattering intensity grows exponentially toward small $q$, and thus, contributions from the smallest $q$ easily dominate the above sum. We found it most useful to fix an arbitrary reference intensity $I_0(q)$ and then measure the similarity of $I_1$ and $I_2$ by the normalized sum
\begin{equation}
	\sum_i \frac{|I_1(q_i) - I_2(q_i)|}{I_0(q_i)}\,. \label{eq:simil}
\end{equation}

Note that the performance of Method 1 can be bolstered by making the number of virtual samples $K$ as large as possible, because a new sample would have a higher chance of finding a very similar counterpart within the database. The workload of determining the best plan for all these samples, however, would significantly increase the computational cost. Therefore, $K$ should be carefully decided by considering the available computational resources and time.

\subsection{\label{sc:m2}Method 2:~Surmise of Future Measurements Through Adaptive Database Search}

This method is based on a comparison of two independent guesses for future measurement outcomes and is conceptually very similar to the approach discussed at the end of section~\ref{sc:setup}. 
Let $Q_{\rm all}$ denote the set of all values of $q$ that can be measured. 
\footnote{At this stage the continuous variable $q$ has to be discretized. In our numerical experiment in section~\ref{sc:exp}, we first fixed the allowed range of $q$ to $[10^{-2.5}, 10^{0.5}]$ and then made a log-uniform grid on it; namely $Q_{\rm all}=\{10^{-2.5+\frac{3}{99}m}\;|\; m=0,1,2,\cdots,99\}$.}
As in Method 1, we again start by creating a database, but it is structurally much simpler than that of \eqref{eq:DD}: it only stores the scattering intensity of $K'$ samples, without measurement plans, as 
\begin{equation}
	\DD' = \Big\{ \{I_k(q)\,|\,q\in Q_{\rm all}\} \Big\}_{k=1}^{K'}\,.
\end{equation}
During an experiment on a new sample, we proceed as follows. Suppose that we obtained data $\II = \{I(q)\,|\,q\in Q_{\rm obs}\}$ as a result of prior measurements. $Q_{\rm obs}$ (with $|Q_{\rm obs}|=n$) denotes the set of $q$ that have been measured. Then, the set of scattering intensities that are currently unknown and will be obtained in future measurements is given by
\begin{equation}
	\II' =\{I(q)\,|\,q\in Q_{\rm all}\setminus Q_{\rm obs}\}\,.
\end{equation}
The core idea of Method 2 is to decide $q_{n+1}$ by inferring $\II'$ in two independent ways. 
\begin{enumerate}
	\item[(A)] The first approach infers $\II'$ in a manner analogous to $k$-nearest-neighbor regression. Specifically, we draw a comparison between $\DD'$ and $\II$ to determine which $I_{k}$ is most similar to the data $\II$. Suppose that $I_a, I_b, I_c$ are the samples in $\DD'$ that most resemble $\II$. Then, we can obtain a guess for $\II'$ by simply taking their average as 
	\begin{equation}
		\big\{[I_a(q)+I_b(q)+I_c(q)]/3\,\big|\,q\in Q_{\rm all}\setminus Q_{\rm obs}\big\}. 
		\label{eq:23}
	\end{equation}
	While we take the average of three samples here, we could vary the number of samples for flexible extraction from $\DD'$. It is also possible to use a more sophisticated machine-learning method in place of a simple mean. 
	\item[(B)] The second approach, which does \emph{not} use the database $\DD'$, proceeds as follows. We first apply the IFT to the data $\II$ and estimate the size distribution function $p(r)$. It is then straightforward to guess $I(q)$ at an arbitrary $q$ by plugging $p(r)$ into \eqref{eq:Idef}. This enables easy estimation of $\II'$. 
\end{enumerate}
The reader might ask why we do not apply simple interpolation techniques to $\II$ to estimate $\II'$. This is an essential question. The key point is that the relation \eqref{eq:Idef} imposes quite strong constraints on the allowed form of $I(q)$. Blindly applying techniques such as cubic splines or Gaussian process (GP) regression to $\II$ turns out to yield an $I(q)$ showing scattering intensity with quite unrealistic behavior. To make a sensible guess at $\II'$, it is mandatory to appropriately incorporate such constraints on $I(q)$. The two methods detailed above address this requirement properly. 

\begin{figure}
	\centering
	\includegraphics[width=.75\columnwidth]{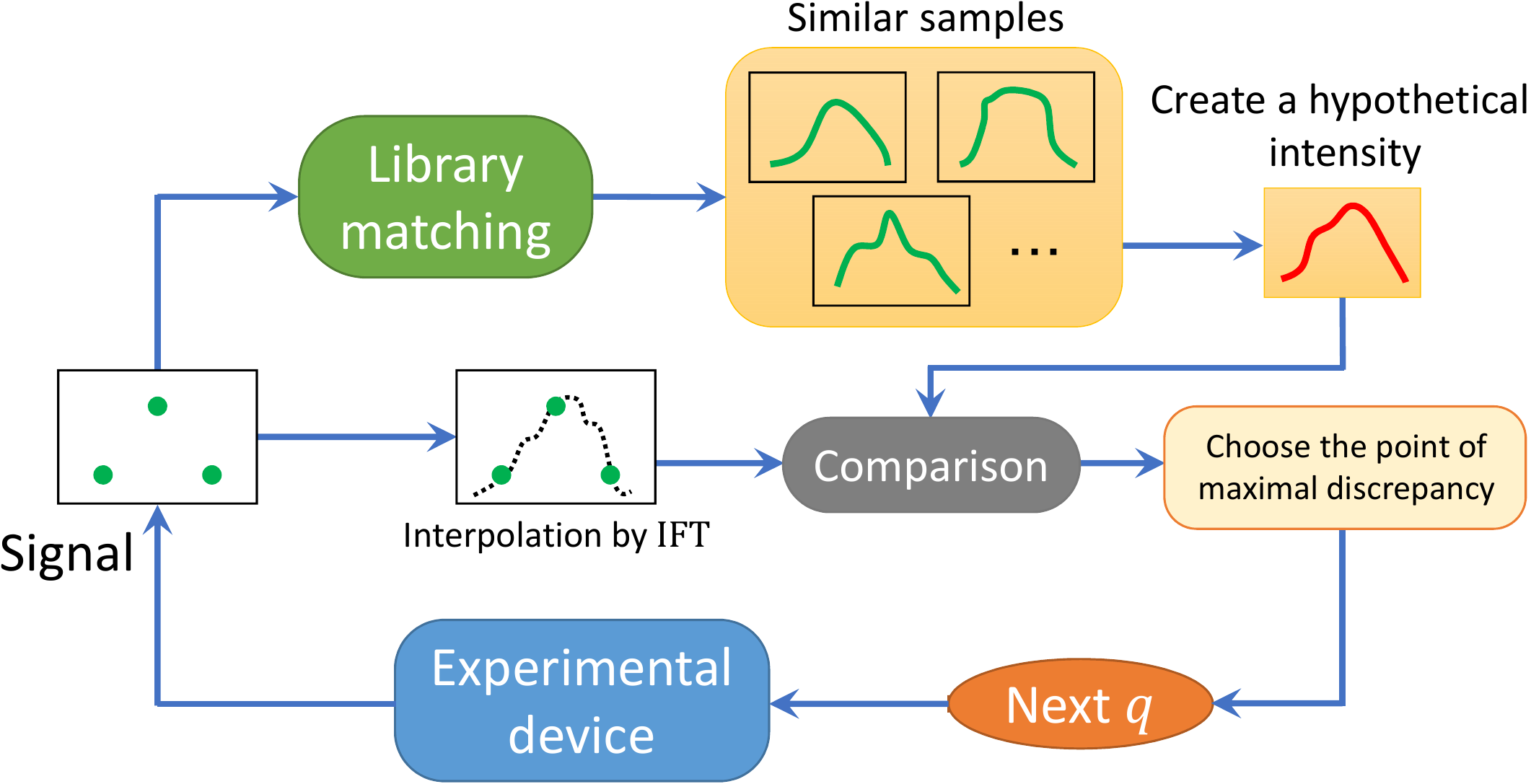}
	\caption{\label{fg:me2}Summary of Method 2.}
\end{figure}

Now, assume that we have obtained two guesses for $\II'$ as $I^{(1)}$ and $I^{(2)}$. Then, we choose the next measurement as 
\begin{equation}
	q_{n+1} = \underset{q\in Q_{\rm all}\setminus Q_{\rm obs}}{\rm arg\,max}
	\frac{\left|I^{(1)}(q)-I^{(2)}(q)\right|}{I_0(q)},
\end{equation}
where $I_0$ is a reference scattering intensity that is set before starting the measurements and held fixed throughout the experiment. After measuring $I(q_{n+1})$, we update $Q_{\rm obs}$ and redo the process of deriving $I^{(1)}$ and $I^{(2)}$. This is iterated until $N$ measurements have been performed. Figure~\ref{fg:me2} schematically summarizes Method 2. (See also Figure~\ref{fg:m2overview} in section~\ref{sc:results} for a supplementary illustration of Method 2.)

Because the validity of Method 2 is less obvious as compared to Method 1, we say a few words about its validity here. Step (B) above obtains a $p(r)$ consistent with the observed data $\II$. In principle, however, there will also be some other distribution functions $p_\alpha(r), p_\beta(r), \cdots$ that are equally consistent with $\II$. This occurs because the data $\II$ is insufficient to uniquely fix the underlying size distribution. What, then, is the best way to discriminate among $p_\alpha(r), p_\beta(r), \cdots$ and $p(r)$? As the distributions themselves are not observable, we have to compare their scattering intensities. In practice, it is difficult if not impossible to construct $p_\alpha(r), p_\beta(r), \cdots$ out of $\II$, so we instead prepare a database of diverse scattering intensities in advance and search for those that resemble $\II$. This is the strategy of Method 2. In evaluating the similarity, we again apply the technique of using \emph{normalized} intensities rather than raw intensities, as in Method 1 (see \eqref{eq:simil}). 

\section{Experiments\label{sc:exp}}
\subsection{Experimental Settings\label{sc:expsettings}}
\begin{figure}
	\centering
	\includegraphics[width=\columnwidth]{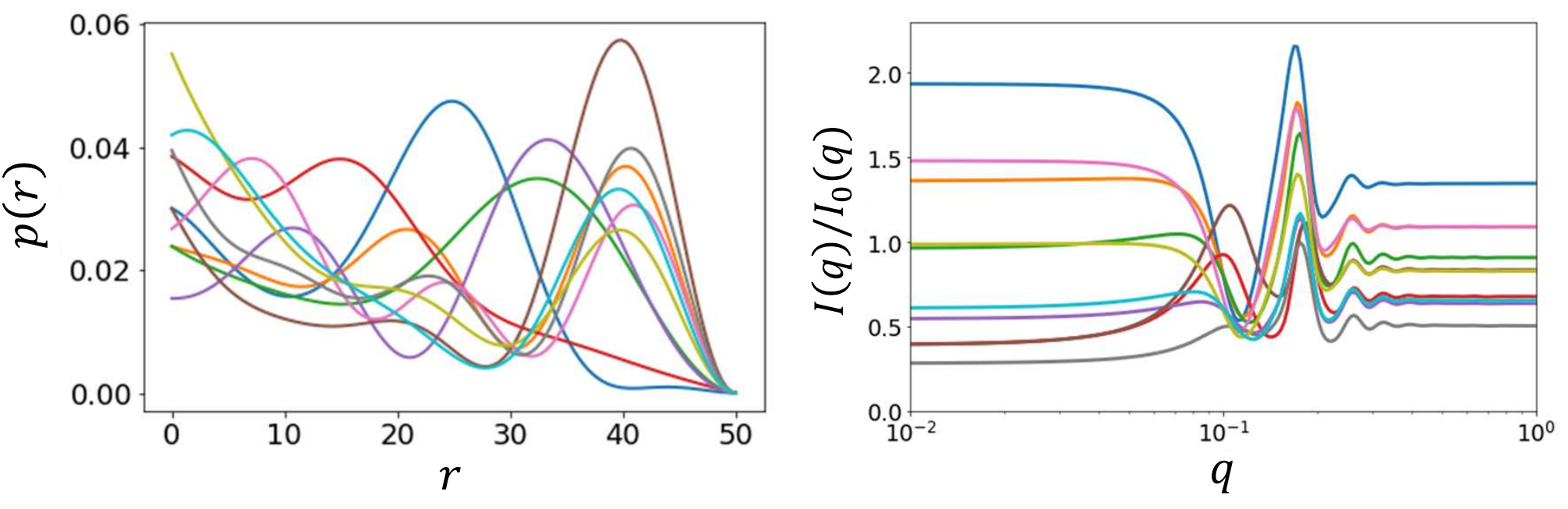}
	\caption{\label{fg:10ex}Ten examples of randomly generated size distributions (left) and their scattering intensities normalized by a reference intensity (right).}
\end{figure}

We tested the effectiveness of Methods 1 and 2 through numerical simulation of SANS experiments on 100 virtual sample materials. We assumed that all size distributions were supported on the range $r\in[0,50]$ and generated 100 random distributions by the formula
\begin{eqnarray}
	p(r) & = & \mathcal{N}\big[ f(r)^2+\beta_1 R + \beta_2 R^2 \big]\,,
	\label{eq:prand}
	\\
	f(r) & = & \alpha_1 \sin R + \alpha_2 \sin 2R + \alpha_3 \sin 3R 
	+ \alpha_4 \sin 4R 
	\notag
	\\
	& & \quad + \alpha_5 \sin 6R + \alpha_6 \sin 8R + \alpha_7 \sin 10R\,,
	\\
	R & = & 1 - r/50\,.
\end{eqnarray}
Here, $\mathcal{N}$ is a normalization factor ensuring that $\int_0^{50}\!\!\text{d}r\,p(r)=1$, $\alpha_{1,\cdots,7}$ are random variables taken from a uniform distribution over $[-1,1]$, and $\beta_{1,2}$ are random variables taken from a uniform distribution over $[0,2]$. By definition, $p(r)$ as generated above is positive definite and vanishes at $r=50$. Figure~\ref{fg:10ex} shows the diversity of 10 examples of $p(r)$ and their normalized scattering intensities. 

We set the range of $q$ to $[10^{-2.5},10^{0.5}]$, so that all measurements occur within this range. The maximal number of measurements $N$ was fixed at 50. We set $K=10^2$ for Method 1 and $K'=10^4$ for Method 2. We found that $K>10^2$ was computationally too expensive for our available resources,
\footnote{In the greedy search stage of Method 1 (see Figure~\ref{fg:tt}), 100 different $q$ were randomly generated as a candidate of $q_{n+1}$ for each $n$. This was repeated for a total of 50 steps until the end of measurement for a given sample. Repeating this procedure for $K=100$ samples amounts to performing the IFT for $100\times50\times100=5\times 10^5$ times in total. Since a single IFT calculation took about 3 seconds on our CPU (Intel  Xeon E5-2650 v3 @ 2.30GHz), the total time required was roughly 400 hours. By parallelizing the calculation on 20 cores we could reduce the time to 20 hours, which was still quite demanding.} 
whereas increasing $K'$ further was feasible but did not bring any performance improvement. 

As a benchmarking test, we compared Methods 1 and 2 with the following three measurement planning methods.
\begin{enumerate}
	\item \emph{Random sampling.} We take 50 points $10^{-2.5+\frac{3}{49}s}\;(s=0,\cdots,49)$ that uniformly cover the allowed range of $q$, $[10^{-2.5},10^{0.5}]$, on a log scale. During an experiment the points are measured in random order without duplication. 
	\item \emph{Sampling based on a Gaussian process (GP).} First, we generate 500 random $p(r)$ according to \eqref{eq:prand} and compute their respective scattering intensities via \eqref{eq:Idef}. The resulting dataset,
	\begin{equation}
		\Big\{\big(\{p_k(r_i)\}_i, \{I_k(q_j)\}_j \big)\Big\}_{k=1}^{500},
	\end{equation}
	is then used to train a GP regression algorithm with an anisotropic Gaussian kernel (we use the \emph{scikit-learn} library for Python \cite{scikit-learn}). After training, the algorithm learns the mapping $\{I_k(q_j)\}_j \to \{p_k(r_i)\}_i$ (see \ref{ap:gpr} for details) and yields the kernel $k(x,y)=\exp(-\sum_j |x_j-y_j|^2/\ell_j^2)$, whose parameters are adjusted through automatic relevance determination \cite{GPbook}. We set $\forall \ell_j=1$ for their initial values. Recalling that the inverse radius $1/\ell_j$ can be interpreted as the importance of $I(q_j)$ for predicting the size distribution, we align the $q_j$ in order of increasing $\ell_j$ and perform measurements in this sequence (i.e., the $q_j$ with the smallest $\ell_j$ is measured first, and the $q_j$ with the largest $\ell_j$ is measured last). As important $q$ are given higher priority, this strategy is expected to rapidly decrease the discrepancy of $p(r)$, even though the order of measurements is entirely fixed and the same for all samples. 
	\item \emph{Sampling based on maximal variance (MV).} This method partly resembles Method 2. First, we prepare a database $\DD'$ through simulation. After every single measurement during an experiment, we obtain the measured data $\II$ and run a similarity search over $\DD'$ to select the $K''$ scattering intensities that most resemble $\II$, where $K''$ is a predefined number. The diversity of these intensities represents the degree of remaining uncertainty under the constraint from $\II$. For the next measurement, we choose the $q$ at which the variance of the $K''$ scattering intensities is greatest, namely,
	\begin{equation}
		q_{n+1} = \underset{q\in Q_{\rm all}\setminus Q_{\rm obs}}{\text{arg\,max}}
		\text{Var}\left[ \big\{ I_k(q)/I_0(q) \big\}_{k=1}^{K''} \right],
	\end{equation}
	where $I_0$ is a reference intensity. By this strategy we can efficiently reduce the uncertainty of the scattering intensity. Because we tested $K''=3,6,12,24$ and found that $K''=12$ performed the best, we use $K''=12$ here.
\end{enumerate} 

During a numerical experiment on a virtual sample, we monitor the performance of the five methods as follows. Let $1\leq n\leq 50$. When the measurement data $\{I(q_j)\}_{j=1}^{n}$ is garnered with any of the five methods, we apply the IFT and obtain the estimated size distribution $\widehat{p}(r)$. We then use both \eqref{eq:l1} and \eqref{eq:KL} to calculate the discrepancy between the estimate and the true distribution and record it. These steps are repeated after every measurement, until $n=50$ is reached. We assume that all methods conduct the first two measurements at $q=10^{-2.5}$ and $q=10^{0.5}$, i.e., at both ends of the allowed range of $q$, meaning that the discrepancy at $n=2$ is exactly the same for all methods. We repeat the entire experimental process for 100 virtual samples and average their performance.

Note that the IFT setup is common to all five methods to enable comparison of their discrepancies on equal footing. Specifically, for the expansion in \eqref{eq:p_exp}, we used 80 base functions defined by
\begin{equation}
	B_k(r) = \cases{
		1 & \text{if} \quad $\frac{5}{8}k \leq r \leq \frac{5}{8}(k+1)$
		\\ 
		0 & \text{otherwise} 
		\\
	}
\end{equation}
for $k=0,\cdots, 79$, which uniformly covers $[0,50]$. We adopted the regularization function $\displaystyle \Omega(a)=\sum_{k=0}^{78}\Big(\ln\frac{a_{k+1}}{a_k}\Big)^2$ and the coefficient $\lambda=10^{-3}$. We made the latter small to avoid bias due to overly strong regularization.

\begin{figure}
	\centering \quad 
	\includegraphics[width=.47\columnwidth]{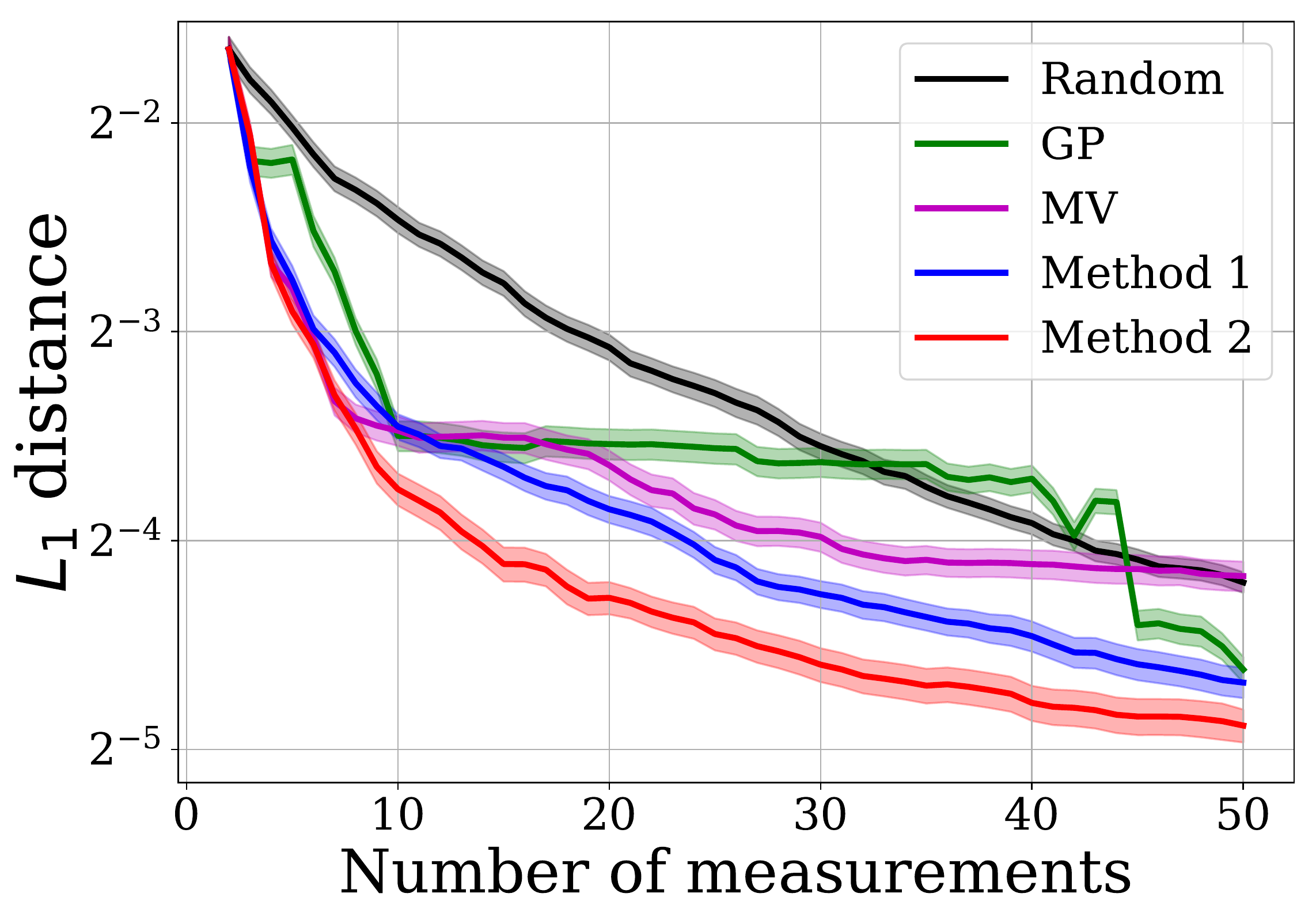}\quad 
	\includegraphics[width=.47\columnwidth]{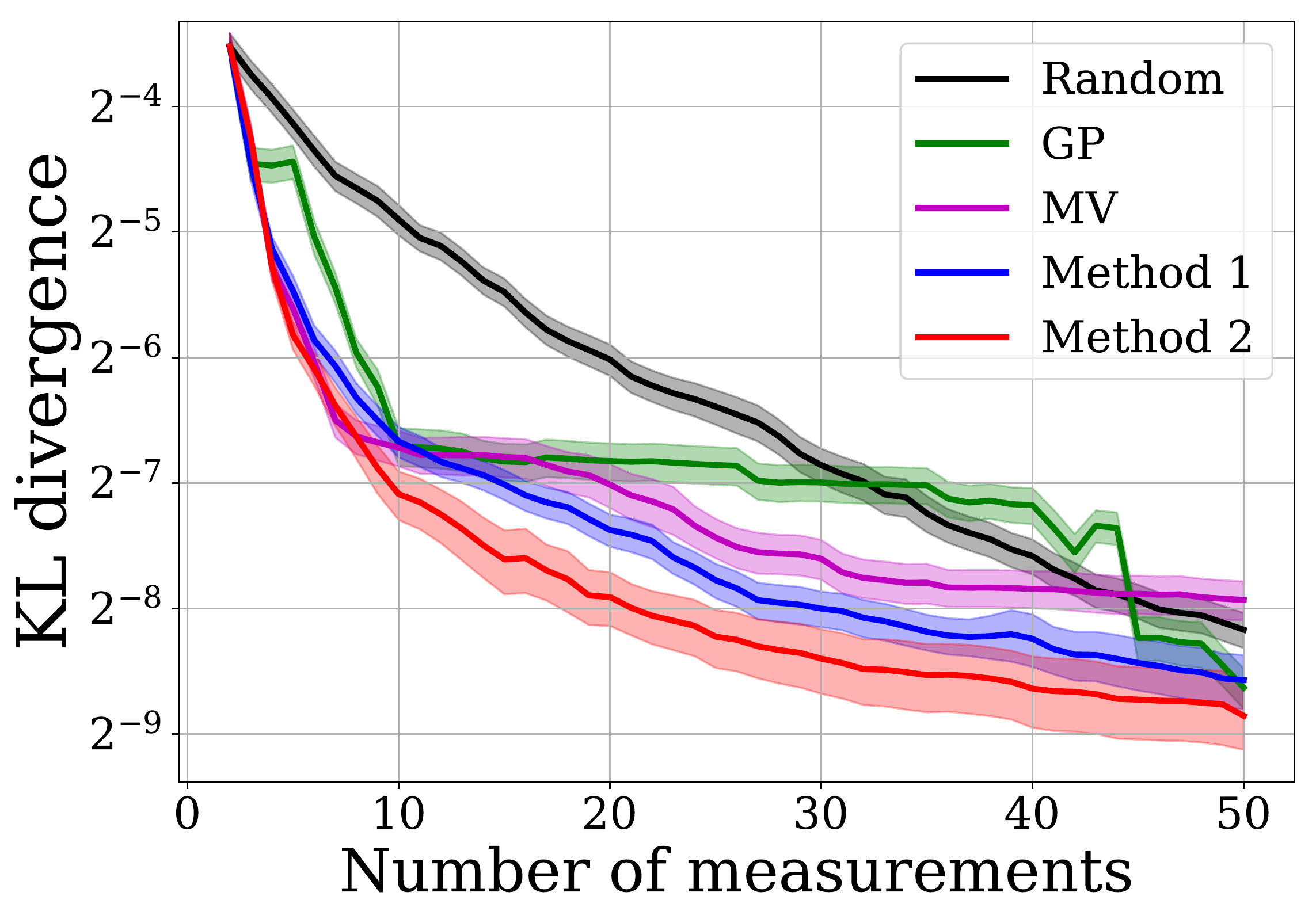}
	\put(-438,142){\bf (a)}
	\put(-218,142){\bf (b)}
	\caption{\label{fg:g}Discrepancy as measured by (a) $L_1$ distance and (b) Kullback--Leibler divergence for the five methods, averaged over 100 virtual samples. ``GP'' and ``MV'' indicate the methods based on the Gaussian process and maximal variance, respectively. Each color band represents one standard deviation.}
\end{figure}

\subsection{Results\label{sc:results}}

We now discuss the results of the benchmarking test. Figure~\ref{fg:g} shows the decrease in the size distribution discrepancy during an experiment, averaged over 100 virtual samples. (Note that the vertical axis of Figure~\ref{fg:g} is on a log scale). Method 2 clearly performed the best. Method 1 was slightly inferior to Method 2 but still outperformed the comparison methods. By contrast, although the GP and MV methods exhibited a rapid initial decrease in the discrepancy, they soon reached a plateau and stagnated. 

\begin{figure}
	\centering \qquad \quad 
	\includegraphics[width=.35\columnwidth]{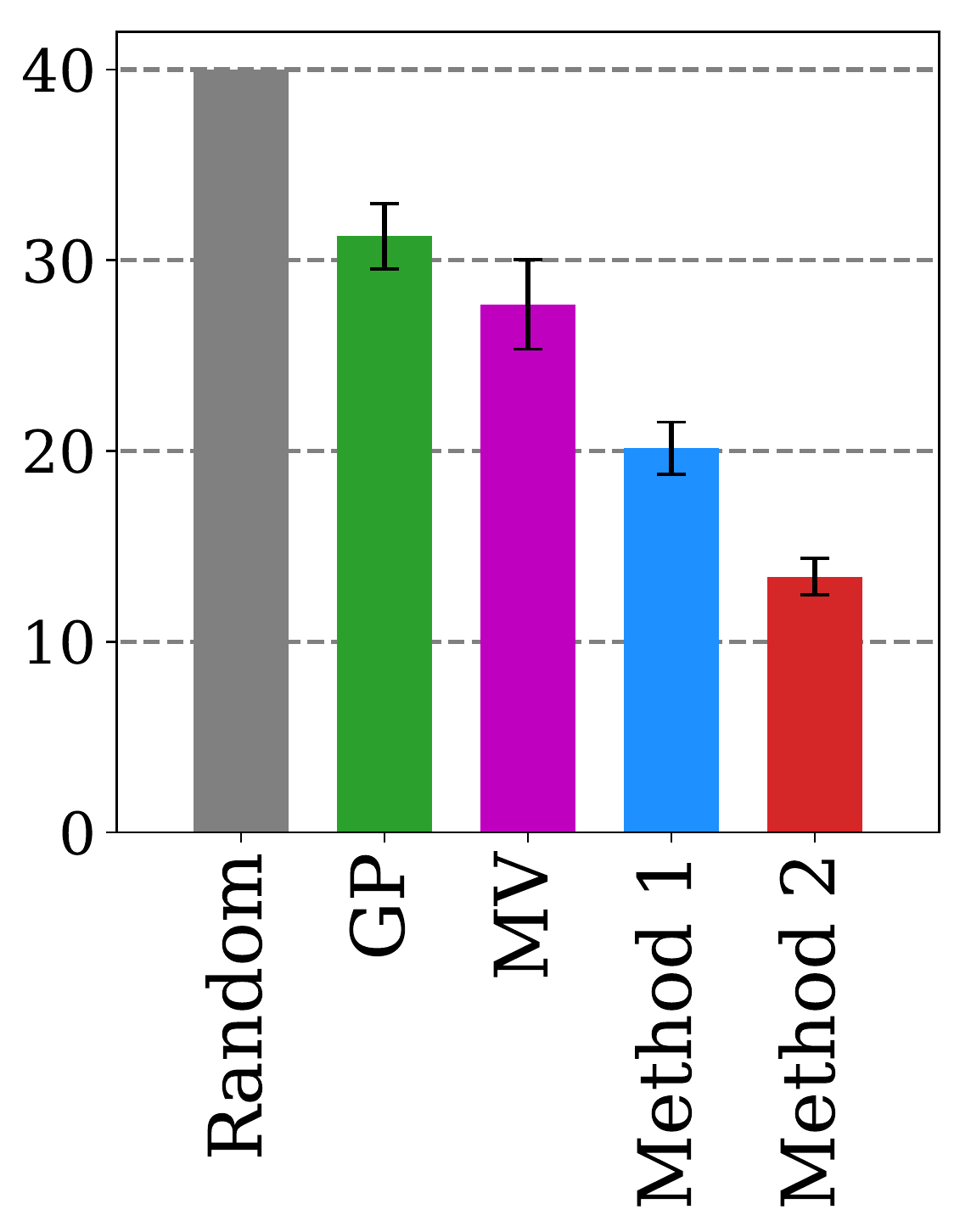}\qquad \qquad 
	\includegraphics[width=.35\columnwidth]{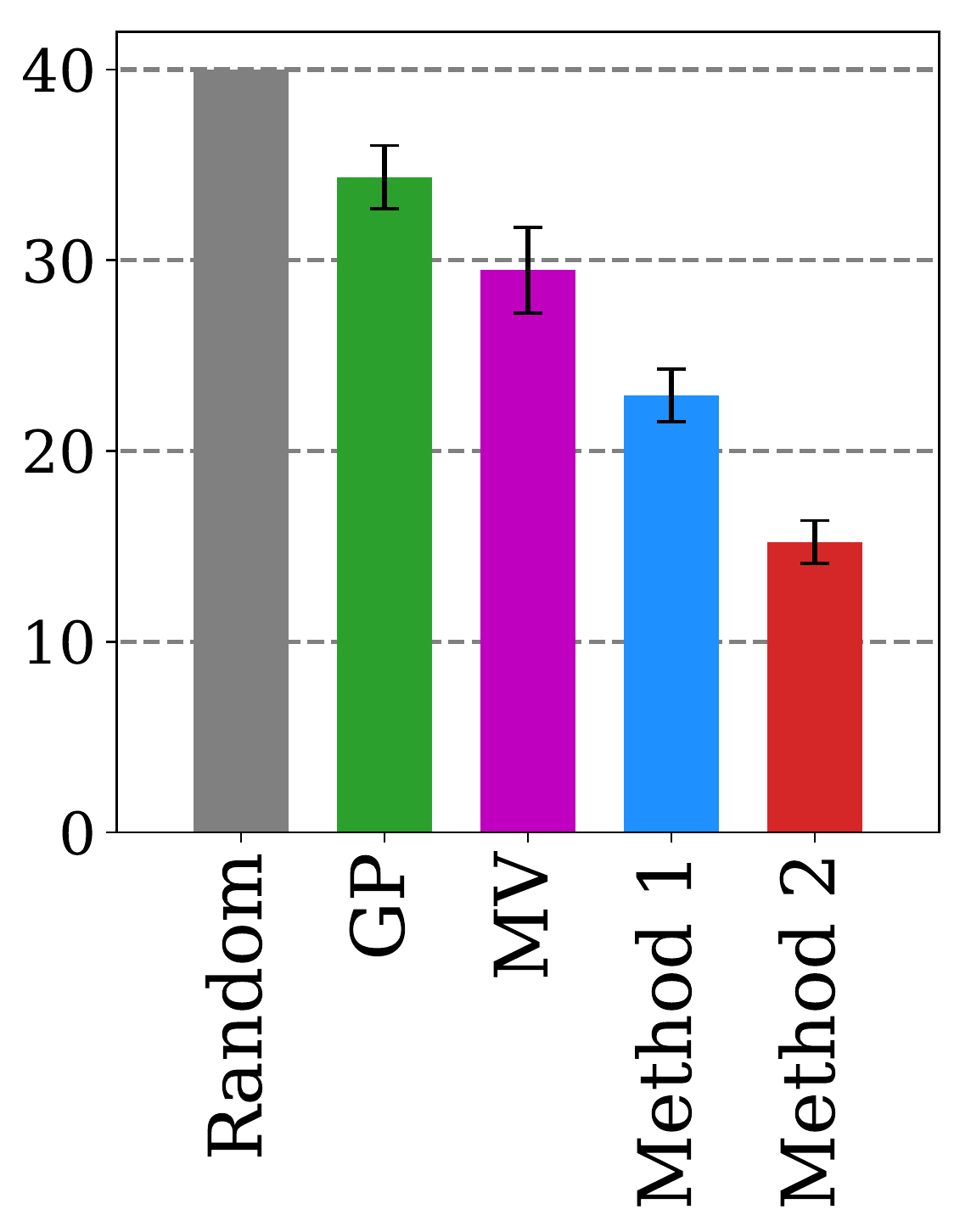}
	\put(-381,187){\bf (a)}
	\put(-389, 87){\rotatebox{90}{$\begin{array}{c}\text{\sf Number~of}\vspace{-5pt}\\\text{\sf measurements}\end{array}$}}
	\put(-176,187){\bf (b)}
	\put(-184, 87){\rotatebox{90}{$\begin{array}{c}\text{\sf Number~of}\vspace{-5pt}\\\text{\sf measurements}\end{array}$}}
	\caption{\label{fg:bar}Average number of measurements required for each method to reach the same discrepancy as that of random sampling with 40 measurements. The discrepancy was measured by (a) $L_1$ distance and (b) Kullback--Leibler divergence. Each error bar represents one standard deviation. The scores are given in Table~\ref{tb:main}.}
\end{figure}

Figure~\ref{fg:bar} and Table~\ref{tb:main} give results for another metric: the number of measurements required for each method to achieve the same discrepancy as random sampling with 40 measurements. Once again Method 2 outperformed the rest, showing roughly $60\sim 65\%$ reduction as compared to random sampling, while Method 1 had roughly 50\% reduction. These results emphasize that adaptive sampling can solve the measurement optimization problem far more efficiently than non-adaptive sampling can. Finally, note that the performance of Method 1 could be improved further by increasing the value of $K$, although this would entail a higher computational cost. 

\begin{table}
	\caption{\label{tb:main}Average number of measurements required for each method to reach the same discrepancy as that of random sampling with 40 measurements. The discrepancy is measured by either $L_1$ distance or KL divergence. The errors represent one standard deviation. The results in bold are the lowest values.}
	\vspace{7pt}
	{\footnotesize 
	\centering
	\begin{tabular}{lrrrrr}  
	\br
	& Random & GP & MV & Method 1 & Method 2 
	\\
	\mr
	\begin{minipage}{.21\textwidth}
	\# of measurements\\($L_1$ distance)
	\end{minipage} & 40 & 31.26 $\pm$ 1.71 & 27.68 $\pm$ 2.34 & 20.14 $\pm$ 1.37 & \textbf{13.41 $\pm$ 0.96}  
	\\
	\mr
	\begin{minipage}{.21\textwidth}
	\# of measurements\\(KL divergence)
	\end{minipage} & 40 & 34.37 $\pm$ 1.66 & 29.48 $\pm$ 2.25 & 22.92 $\pm$ 1.38 & \textbf{15.23 $\pm$ 1.12} 
	\\
	\br
	\end{tabular}
	}
\end{table}

\begin{figure}
	\centering 
	\includegraphics[width=\columnwidth]{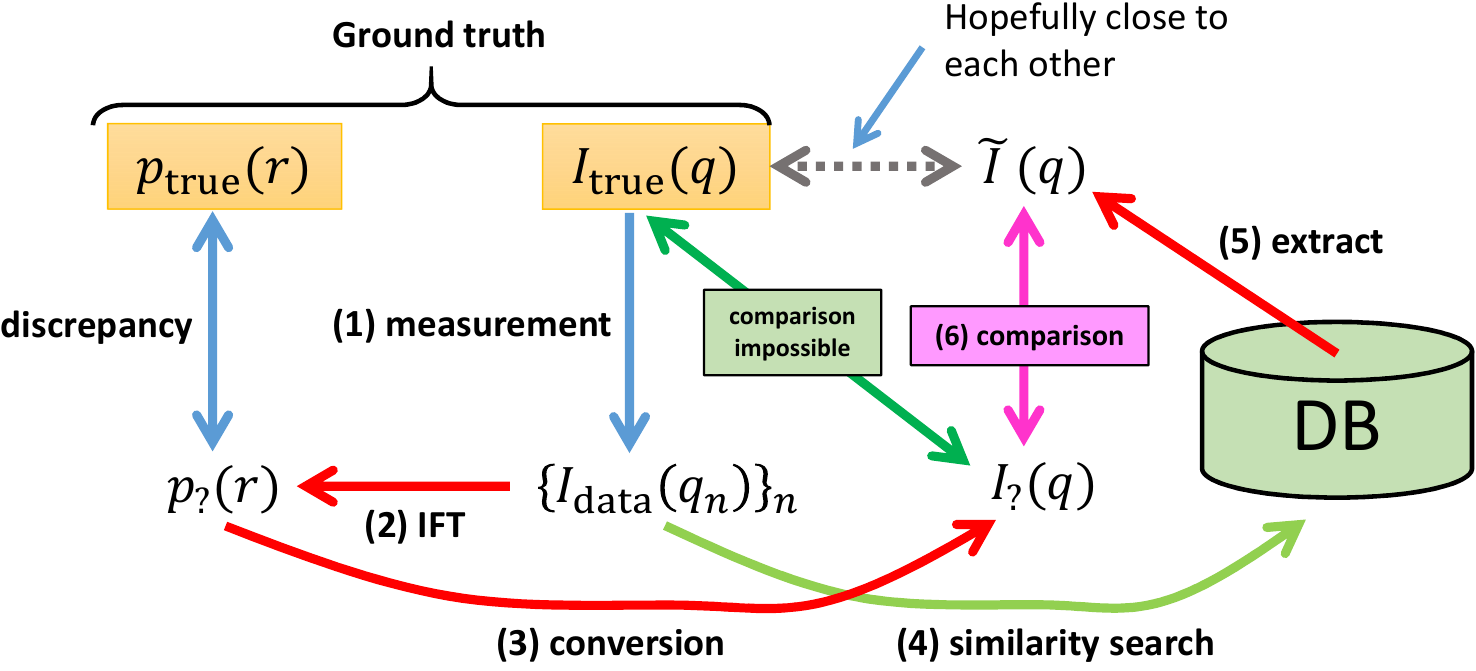}
	\caption{\label{fg:m2overview}Dissection of Method 2. One cycle of the procedure starts from (1) measurement and ends by (6) comparison. See the main text for more details.}
\end{figure}

Why does Method 2 perform better than Method 1 and all other methods? The reason is probably because Method 2 tries to correct the current estimate and push it towards the ground truth far more directly than all the other methods do. For illustration, we display the gist of Method 2 in Figure~\ref{fg:m2overview}. We start off with measurements of the scattering intensity, collect data and then apply the IFT (cf.~section~\ref{sc:sansdefs}) to infer the size distribution $p_{?}(r)$. Ideally the next measurement should be designed so as to maximally decrease the discrepancy between the true distribution $p_{\rm true}(r)$ and our estimate $p_?(r)$, but the discrepancy cannot be calculated without knowing $p_{\rm true}$. So we next turn our focus to comparing scattering intensities. It is straightforward to convert $p_?(r)$ to the scattering intensity $I_?(q)$ via \eqref{eq:Idef}. However, we again stumble on the difficulty that $I_?(q)$ and the true intensity $I_{\rm true}(q)$ cannot be directly compared since the latter is known only after sufficiently many measurements have been performed. Thus we construct a \textit{surrogate} of $I_{\rm true}(q)$ by utilizing a database of intensities, which is denoted as $\widetilde{I}(q)$ in Figure~\ref{fg:m2overview} (recall \eqref{eq:23}). The next measurement is designed so as to maximally decrease the discrepancy between $I_?(q)$ and $\widetilde{I}(q)$. One can envisage that it would naturally lead to large reduction of discrepancy between $p_?(r)$ and $p_{\rm true}(r)$ as well. The fact that Method 2 empirically performs best underscores the effectiveness of this surrogate-building idea. The success of Method 2 hinges on the close proximity of $\widetilde{I}(q)$ to $I_{\rm true}(q)$, which was warranted through two ingredients. First, the database of size $K'=10^4$ was large enough for this method to work. Second, the similarity of intensities was measured through the \emph{normalized} distance as in \eqref{eq:simil}, which proved far more effective than just comparing raw intensities. We think these were keys to the success of Method 2.

\begin{figure}
	\centering 
	\includegraphics[width=.8\columnwidth]{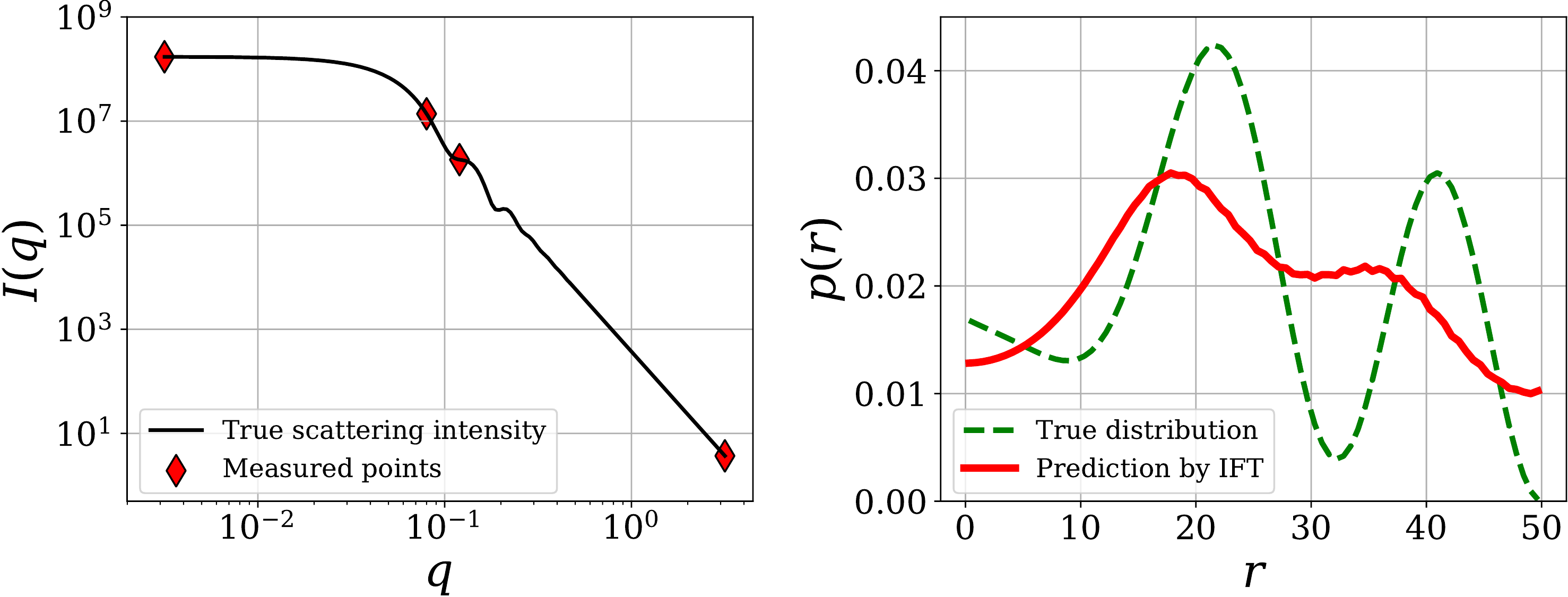}
	\vspace{2mm}\\
	\includegraphics[width=.8\columnwidth]{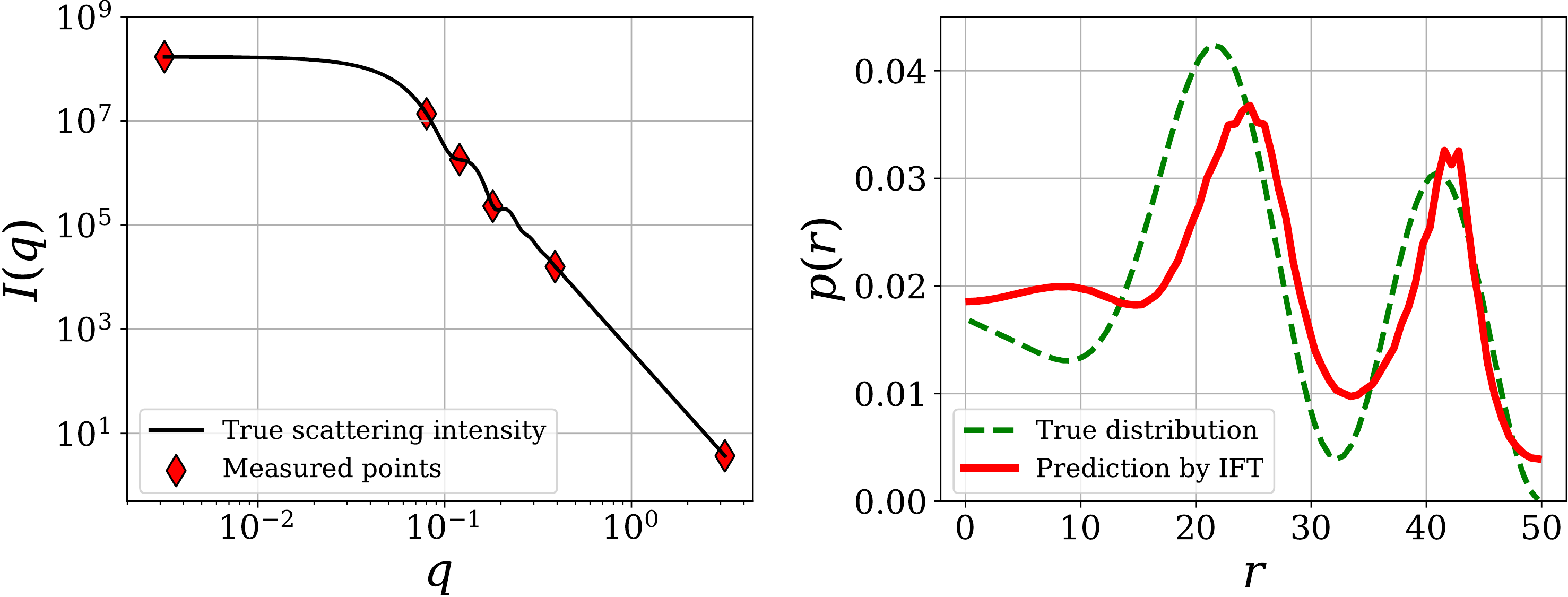}
	\vspace{2mm}\\
	\includegraphics[width=.8\columnwidth]{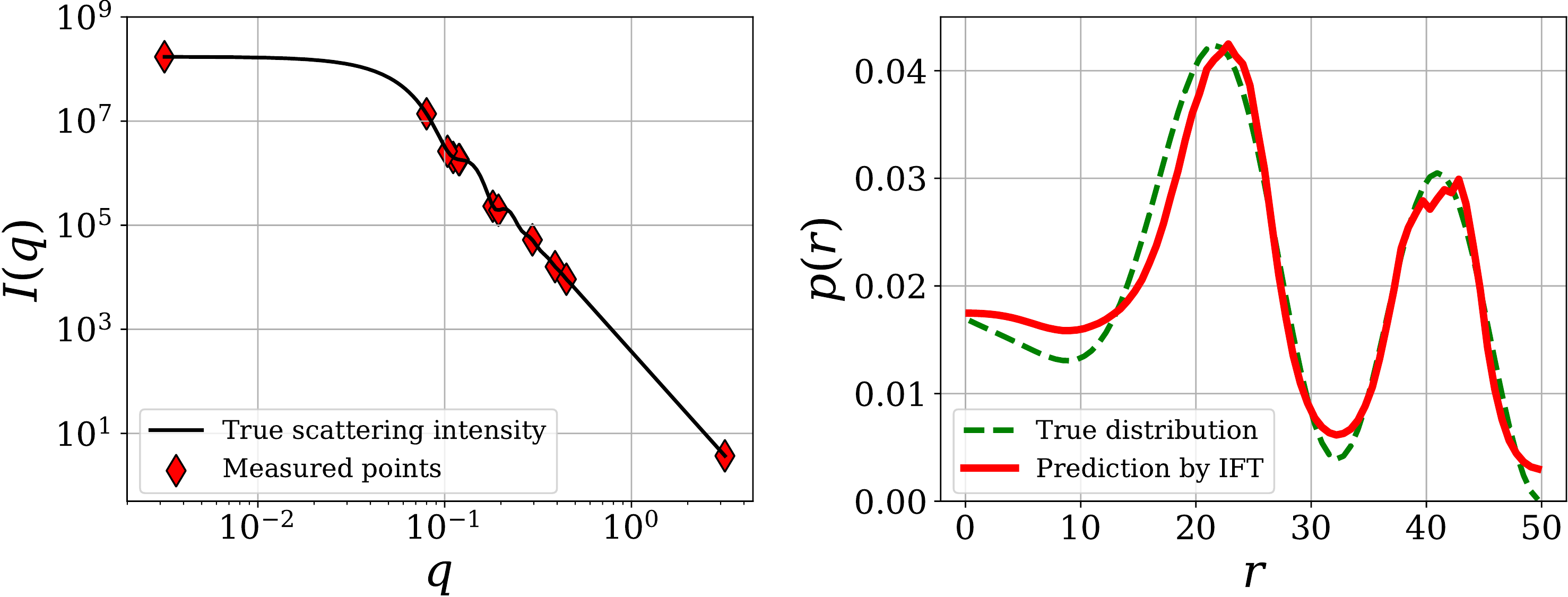}~
	\caption{\label{fg:4612}Example of a measurement process with Method 2. The left column is the scattering intensity and the right column is the associated size distribution (both the ground truth and the predicted one). The number of sampled points is 4 (top row), 6 (middle row) and 12 (bottom row), respectively.}
\end{figure}

In Figure~\ref{fg:4612} we show the process of data sampling with Method 2 for one sample material. It is observed that the accuracy of the IFT prediction for the size distribution improves rapidly as the number of measured points increases. 

As seen above, both Method 1 and 2 make the measurement process more efficient than random sampling, but one may ask how much time was saved overall if one took into account both the time needed to prepare the databases $\DD$ and $\DD'$, and the time for similarity search during the adaptive sampling process. As already remarked in the footnote in section~\ref{sc:expsettings},  the preparation of $\DD$ can take $O(10)$ hours, which is indeed quite huge time investment. However it should be underlined that, once $\DD$ is prepared, one can repeatedly make use of it for measurements of various sample materials for arbitrarily many times; it will eventually pay off if $\DD$ was used to expedite sufficiently many experiments. By contrast the preparation of $\DD'$ is computationally much less expensive; it took us less than 1 hour to prepare $\DD'$ of size $K'=10^4$, so it is not worrisome. On the other hand, the time for similarity search during sampling is essentially negligible compared to the time for preparation of the databases. In our numerical experiment, it only took a few  seconds to finish a search per measurement. (Method 2 needs slightly more time than Method 1 due to the IFT calculation.) Of course this may change in other situations, where for some reason a huge database of size $\gg 10^4$ is needed; then the time for searching may not be negligible and a separate consideration is required.

\section{Conclusions\label{sc:conc}}
To design a new useful material generally requires a repetition of laborious experiments to accurately measure materials properties, and there is high demand for developing tools to minimize the time, cost, and effort involved in sequential high-precision experiments. In this work, we have proposed a new approach to implementing an adaptive policy for multi-step decision making, taking a small-angle neutron scattering (SANS) experiment in materials science as a proving ground. Through numerical experiments we demonstrated that the two methods proposed in this paper can speed up a SANS experiment by a factor of 2--3 as compared to random sampling without prior planning. We expect that our methods can be used to speed up a small-angle X-ray scattering (SAXS) experiment as well. Extending our scheme to SANS experiments for non-spherical scatterers is an important future direction of research. Future work will also include addressing general sequential optimization problems beyond SANS experiments, as well as investigating the effect of noisy observation on the proposed methods. 

In this paper we exclusively relied on numerical simulations to assess the proposed methods; however, testing them in real SANS/SAXS experiments is probably of utmost importance for experimental physicists. We defer this task to future work. 

\ack
We thank the anonymous referees for valuable comments which contributed to the significant improvement of the paper.

\appendix
\section{Training stage of the Gaussian process regression\label{ap:gpr}}

This appendix includes some supplementary remarks on the sampling method based on the Gaussian process (GP) regression introduced in section~\ref{sc:expsettings}. 

Assume that a training dataset $\big\{\mathbf{x}^{(i)}, \mathbf{y}^{(i)}\big\}_{i=1}^{N}$ is given, where $\mathbf{x}^{(i)}$ is a $p$-dimensional input vector and $\mathbf{y}^{(i)}$ a $q$-dimensional output vector. In general, the goal of regression is to construct a model that can predict the output $\mathbf{y}$ at a new input point $\mathbf{x}$. In the GP regression  \cite{GPbook}, the result of model training is given in terms of a kernel function $k(\cdot,\cdot):\mathbb{R}^p\times \mathbb{R}^p \to \mathbb{R}$ as $\mathbf{y}=\sum_{i=1}^{N}\bm{\alpha}^{(i)}k(\mathbf{x}^{(i)}, \mathbf{x})$, where $\bm{\alpha}^{(i)}\in\mathbb{R}^q$ are coefficients that are learned through training and  $k(\mathbf{x}^{(i)}, \mathbf{x})$ essentially represents the proximity of $\mathbf{x}^{(i)}$ to $\mathbf{x}$. This way a smooth multi-dimensional mapping $\mathbb{R}^p\to\mathbb{R}^q$ is learned. Not only that, the GP regression can also predict the range of uncertainty of a prediction, but it is not our focus here.  

Let us return to the problem in section~\ref{sc:expsettings}. The input of regression $\{\mathbf{x}^{(i)}\}_{i=1}^{N}$ is now given by 
$\big\{\bm{I}^{(i)}\big\}_{i=1}^{500}$ where
\begin{equation}
	\bm{I}^{(i)} = \bigg(\ln I_i(q_1), \ln I_i(q_2), \cdots, \ln I_i(q_{100})\bigg) \in \mathbb{R}^{100}
\end{equation}
is the logarithm of the scattering intensity of the $i$-th sample material evaluated at $\displaystyle q_k \equiv 10^{-2.5+\frac{3}{99}(k-1)}$, which uniformly covers the range $10^{-2.5}\leq q \leq 10^{0.5}$ on a log scale.  
We took the log of the intensity to numerically stabilize the learning algorithm. 
On the other hand, the output of regression $\{\mathbf{y}^{(i)}\}_{i=1}^{N}$ is given by $\big\{\bm{p}^{(i)}\big\}_{i=1}^{500}$ where
\begin{equation}
	\bm{p}^{(i)} = \Big(p_i(r_1), p_i(r_2), \cdots, p_i(r_{80})\Big) 
	\in \mathbb{R}^{80}
\end{equation}
is the size distribution function of the $i$-th sample evaluated at $r_k\equiv 50\times\frac{k-1/2}{80}$, which uniformly covers the range $0\leq r \leq 50$.  
Through a training process the GP algorithm constructs a model that learns a highly nonlinear mapping from $\mathbb{R}^{100}$ to $\mathbb{R}^{80}$. During training the algorithm judges which components of $\bm{I}^{(i)}$ are more important than others for accurate prediction of $\bm{p}^{(i)}$. It eventually allows us to learn which part of the scattering intensity controls most of the size distribution. This information is used to assign appropriate priority of measurement to each $q_k$.

\section*{References}
\bibliography{draft_v4_after_referee_review.bbl}
\end{document}